\documentclass[prb,twocolumn,floatfix,raggedbottom,raggedfooter,altaffilletter]{revtex4}
\usepackage[tbtags]{amsmath}
\usepackage{amssymb}
\usepackage{bm}
\usepackage{graphicx}
\usepackage{dcolumn}
\newcommand{\nodag}{\rule[-0.65ex]{0pt}{1.35ex}}
\newcommand{\half}{\frac{1}{2}}

\newcommand{\thalf}{\mbox{$\textstyle\half$}}

\begin{document}

\title{Absence of Kondo lattice coherence effects in Ce$_{0.6}$La$_{0.4}$Pb$_3$:
A magnetic-field study}

\author{Richard Pietri}
\altaffiliation[Present address: ]{Alien Technology, 18220 Butterfield Blvd.,
Morgan Hill, CA 95037}
\affiliation{Department of Physics, University of
Florida, Gainesville, Florida 32611-8440}

\author{Costel R.\ Rotundu}
\affiliation{Department of Physics, University of Florida, Gainesville,
FL 32611--8440}

\author{Bohdan Andraka}
\affiliation{Department of Physics, University of Florida, Gainesville,
FL 32611--8440}

\author{Bryan C.\ Daniels}
\altaffiliation[Permanent address: ]{Department of Physics
and Astronomy, Ohio Wesleyan University, Delaware, OH 33015}
\affiliation{Department of Physics, University of Florida, Gainesville,
FL 32611--8440}

\author{Kevin Ingersent}
\email[Author to whom correspondence should be addressed; electronic
mail: ]{ingersent@phys.ufl.edu}
\affiliation{Department of Physics, University of Florida, Gainesville,
FL 32611--8440}


\begin{abstract}
The specific heat of polycrystalline Ce$_{0.6}$La$_{0.4}$Pb$_3$ has been
measured in magnetic fields ranging from 0 to 14\,T.
After subtraction of a lattice contribution, the specific heat between
1\,K and 10\,K is well described by the $S=\half$ single-impurity Kondo model
with just one adjustable parameter: the zero-field Kondo temperature.
In particular, the variation in the temperature and the height of the peak in
$C$ vs $T$ is captured with good accuracy.
This fit suggests that lattice coherence effects play no significant role in
the magnetic-field response of this concentrated Kondo system.
\end{abstract}

\date{28 October 2004}

\maketitle


\section{INTRODUCTION}

Lately, there has been renewed interest in the subject of the Kondo lattice
and its relation to the single-impurity Kondo model. Nakatsuji
\textit{et al.}\cite{Nakatsuji:04} have proposed a two-fluid model that
 thermodynamic and transport properties of Ce$_x$La$_{1-x}$CoIn$_5$
by the superposition of a single-impurity part and a coherent heavy-fermion
liquid part.
It has been shown for the concentrated Ce alloys ($x > 0.5$) that only 10\%
of the low-temperature specific heat corresponds to the single-impurity part,
and that this part can be described by the same Kondo temperature $T_K$ for
all concentrations $x$.

This remarkable result, which suggests that the specific heat is essentially a
coherent lattice property, requires reexamination of previous investigations
concluding that single-impurity physics accounts rather well for the
thermodynamic properties of a number of heavy fermions. Of particular note is
the alloy series Ce$_x$La$_{1-x}$Pb$_3$, for which the zero-field specific
heat scales with Ce concentration for $0<x\le 0.6$, and the
specific heat per Ce is accounted for quantitatively by the $S=\half$
single-impurity Kondo model.\cite{Lin:87}
The absence of coherence effects in Ce$_x$La$_{1-x}$Pb$_3$ is surprising given
that the pure compound CePb$_3$ orders antiferromagnetically at $T_N = 1.1\,$K,
pointing to the presence of significant inter-ion correlations.

In order to provide a more rigorous test of the single-impurity picture in
Ce$_x$La$_{1-x}$Pb$_3$, we have measured the specific heat of polycrystalline
Ce$_{0.6}$La$_{0.4}$Pb$_3$ in magnetic fields ranging from 0 to 14\,T.
The $x=0.6$ concentration was chosen to satisfy two criteria: the system
should be sufficiently concentrated and it should not order magnetically
at any temperature.
After subtraction of a lattice contribution, we find that the specific heat
between 1\,K and 10 \,K is well described by the $S=\half$ single-impurity
Kondo model with just one adjustable parameter: the zero-field Kondo
temperature, $T_K = 2.6 \pm 0.2 \,$K.
In particular, the variation in the temperature and the height of the peak in
$C$ vs $T$ is captured with good accuracy.
This fit is non-trivial, given that the impurity has a field-dependent $g$
factor arising from field-induced mixing of Ce crystalline electric field
levels.

\section{EXPERIMENT}

Two polycrystalline samples of Ce$_{0.6}$La$_{0.4}$Pb$_3$ were synthesized
independently in an arc melter, using highest available grade elements (Ce
and La from Ames National Laboratory, Pb 6N from AESAR Johnson Matthey).
Because of the high vapor pressure of Pb, the starting material had
additional Pb to compensate for vapor losses.
The starting composition for sample 1 had 3\% more Pb than indicated
by stoichiometry, while that for sample 2 had an additional 2\%.
Each sample was repeatedly remelted to improve the homogeneity.
After each remelting, the sample mass was compared to that expected for the
stoichiometric material under the assumption that there were no vapor losses
of Ce and La at the low arc current used.
The process was repeated until the final stoichiometry (assuming no
loss of Ce and La) was Ce$_{0.6}$La$_{0.4}$Pb$_{3.00\pm 0.01}$.
Each sample was then annealed for two weeks at 600\,$^{\circ}$C in the
presence of additional free lead to minimize further Pb losses from
the sample. (No such losses were detected.)

\begin{figure}[b]
\includegraphics{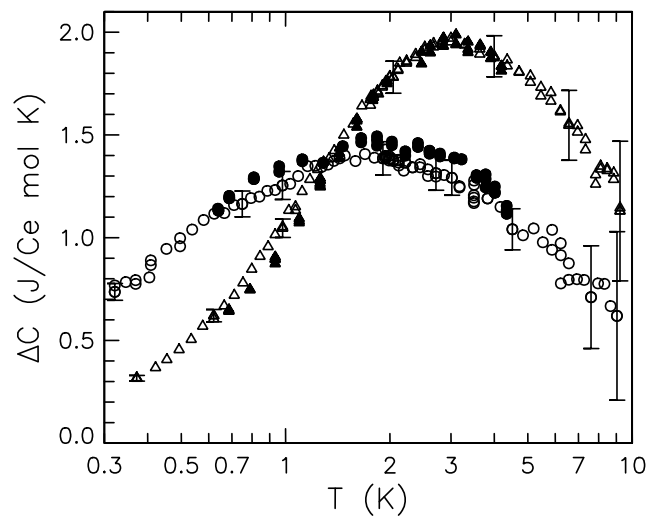}
\vspace*{-2ex}
\caption{\label{fig:both_samples}%
Excess specific heat $\Delta C$ vs temperature $T$ for sample 1 (open symbols)
and sample 2 (filled symbols) at magnetic fields $H=0$ (circles) and $H=10$\,T
(triangles). Error bars are shown for a few representative points.
}
\end{figure}

The specific heat $C$ of sample 1 was measured by the thermal relaxation
method over temperatures $T$ between 0.4\,K and 10\,K in magnetic fields
$H=0$, 5, 8, 10, and 14\,T.
Sample 2 was measured between 0.7\,K and 4.2\,K at zero field and 10\,T
to provide a basis for estimating the likely degree of sample dependence
in the data.

\begin{figure}[b]
\centering
\includegraphics{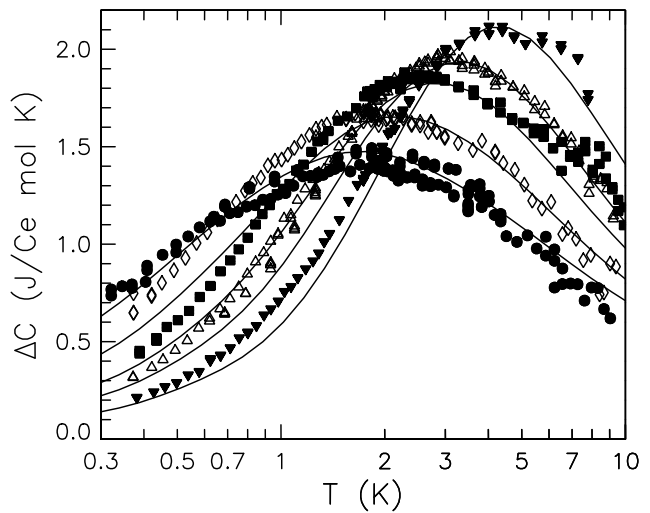}
\vspace*{-2ex}
\caption{\label{fig:compare}%
Excess specific heat $\Delta C$ vs temperature $T$ in fields of 0\,T
({\large$\bullet$}), 5\,T ({\small$\diamondsuit$}), 8\,T
({\small$\blacksquare$}), 10\,T ($\triangle$),
and 14\,T ($\blacktriangledown$).
Data for samples 1 and 2 are combined.
The lines show $C_{\mathrm{imp}}$ for the single-impurity Kondo model
[Eq.~\protect\eqref{H_K}] calculated for the experimental $H$ values,
with $g_i$ set to the corresponding $\langle g_i\rangle$ listed in
Table~\protect\ref{table}.
}
\end{figure}

Figures~\ref{fig:both_samples} and~\ref{fig:compare} plot the excess
specific heat $\Delta C = C - C_{\mathrm{lat}}$ normalized per mole of Ce.
Here, $C_{\mathrm{lat}}$ is the lattice (phonon) contribution, estimated
from the data of Lin \textit{et al.}\cite{Lin:87}
Figure~\ref{fig:both_samples} also shows error bars for a few representative
points.
The uncertainties reflect possible errors both in the measured total specific
heat and in the lattice correction.
By 10\,K, $C_{\mathrm{lat}}$ makes up roughly 90\% of $C$, so the uncertainty
in $\Delta C$ is particularly large at the upper end of the measured
temperature range.

\section{THEORY}

We modeled the experimental data using the spin-\thalf\ Kondo
impurity model described by the Hamiltonian
\begin{eqnarray}
\label{H_K}
\hat{H}_K &=& \sum_{\mathbf{k},\sigma} (\epsilon_{\mathbf{k}} +
\sigma g_c \mu_B H) \, c^{\dagger}_{\mathbf{k}\sigma}
c^{\nodag}_{\mathbf{k}\sigma} + g_i \mu_B H S_z
\nonumber \\
&& + \, J \mathbf{S} \cdot
\sum_{\mathbf{k},\sigma} \sum_{\mathbf{k}',\sigma'}
c^{\dagger}_{\mathbf{k}\sigma} \thalf \bm{\sigma}_{\sigma\sigma'}
c^{\nodag}_{\mathbf{k}'\sigma'} .
\end{eqnarray}
Here $\epsilon_{\mathbf{k}}$ describes the conduction-band dispersion,
$\sigma = \pm\half$ labels the conduction-electron spin projection along
the direction of the magnetic field, $\mu_B$ is the Bohr magneton,
$g_c$ and $g_i$ are the conduction-band and impurity $g$ factors,
respectively, $J$ is the Kondo exchange, $\mathbf{S}$ is the impurity
spin operator, and $\sigma^j_{\sigma\sigma'}$ ($j = 1$, 2, 3) are the
Pauli matrices.

At zero field, the model has a single low-energy scale\cite{Wilson:75}
$k_B T_K \approx \epsilon_F \exp[-1/\rho(\epsilon_F) J]$, where
$k_B$ is Boltzmann's constant, and
$\rho(\epsilon_F)$ is the density of states at the Fermi energy $\epsilon_F$.
Since the quantities entering the Kondo temperature $T_K$ were not measured
experimentally, we calculated $C_{\mathrm{imp}}$, the impurity contribution
to the heat capacity, using the numerical renormalization group (NRG) method
\cite{Wilson:75,Oliveira:94} for an arbitrary choice $\rho(\epsilon_F)J
= 0.2$ and then fitted the temperature scale of the data for
Ce$_{0.6}$La$_{0.4}$Pb$_3$, focusing particularly on the region around the
peak in $C$ (which occurs in both samples at $T \approx 1.8 \pm 0.2$\,K).
This process yielded a value $T_K = 2.6\pm 0.2\,K$, some 20\% lower than
that obtained by Lin \textit{et al.}\cite{Lin:87}
We have no explanation for this discrepancy.

In magnetic fields, it is also necessary to know $g_c$ and $g_i$.
The calculated $C_{\mathrm{imp}}$ is insensitive to the value of
the conduction-band $g$ factor, which we took to be $g_c=2$.
The impurity $g$ factor $g_i$ is deduced by mapping the lowest pair of energy
levels of Ce$^{3+}$ onto an effective spin-\thalf\ degree of freedom,
as described in the remainder of this section.

\begin{table}[b]
\begin{ruledtabular}
\begin{tabular}{@{\extracolsep{1em}}rccccc}
$H$(T) & $\tilde{T}_{\mathrm{CEF}}$(K) &
$\mathrm{max}\,g_i$ & $\mathrm{min}\,g_i$ &
$\langle g_i\rangle$ & $\sigma(g_i)$ \\ \hline
 0\rule{0.5em}{0ex} & 72 & 1.429 & 1.429 & 1.429 & 0.000 \\
 5\rule{0.5em}{0ex} & 65 & 1.431 & 1.412 & 1.423 & 0.005 \\
 8\rule{0.5em}{0ex} & 61 & 1.434 & 1.387 & 1.415 & 0.012 \\
10\rule{0.5em}{0ex} & 59 & 1.436 & 1.364 & 1.407 & 0.019 \\
14\rule{0.5em}{0ex} & 55 & 1.443 & 1.302 & 1.387 & 0.037
\end{tabular}
\end{ruledtabular}
\caption{\label{table}%
Properties of Ce$^{3+}$ in CePb$_3$ cubic crystalline electric fields with
an applied magnetic field of magnitude $H$.
$k_B \tilde{T}_{\mathrm{CEF}}$ is the minimum over all field orientations of
the energy gap between the second and third levels.
$\mathrm{max}\,g_i$ and $\mathrm{min}\,g_i$ are the maximum and minimum
values over all field orientations of the effective impurity $g$ factor
deduced from the splitting between the first and second energy levels.
$\langle g_i\rangle$ and $\sigma(g_i)$ are the mean and standard deviation
of $g_i$, respectively.}
\end{table}

In the cubic crystalline electric field (CEF) environment of
Ce$_x$La$_{1-x}$Pb$_3$, the six $J = 5/2$ atomic levels of atomic Ce$^{3+}$,
$\{ |m_J\rangle \}$, split into a
$\Gamma_7$ doublet and a $\Gamma_8$:quartet\cite{Lea:62}
\begin{eqnarray}
|\Gamma_7,\pm\rangle &=& \textstyle
\sqrt{\frac{1}{6}} \, |\pm\frac{5}{2}\rangle -
\sqrt{\frac{5}{6}} \, |\mp\frac{3}{2}\rangle, \nonumber \\
|\Gamma_8,1,\pm\rangle &=& \textstyle
\sqrt{\frac{5}{6}} \, |\pm\frac{5}{2}\rangle +
\sqrt{\frac{1}{6}} \, |\mp\frac{3}{2}\rangle, \\
|\Gamma_8,2,\pm\rangle &=& \textstyle
|\pm\frac{1}{2}\rangle. \nonumber
\end{eqnarray}
The preponderance of experimental evidence%
\cite{Lethuillier:76,Nikl:87,Renker:87} indicates that CePb$_3$ has a
$\Gamma_7$ ground state.\cite{Durkop:86}
The $\Gamma_7$-$\Gamma_8$ splitting temperature $T_{\mathrm{CEF}}$ has been
variously estimated from the magnetic susceptibility\cite{Lethuillier:76}
to be 67\,K, from the elastic constants \cite{Nikl:87} to be 76\,K, and from
inelastic neutron scattering to be 67\,K (Ref.~\onlinecite{Renker:87}) and
72\,K (Ref.~\onlinecite{Vettier:86}).
It is probable that the CEF scheme is affected only weakly by substitution of
La for some Ce atoms since the immediate environment of each remaining Ce is
unaffected, so we set $T_{\mathrm{CEF}}=72$\,K in our calculations.

In a magnetic field $\mathbf{H}$, the CEF states are mixed by the Zeeman
interaction.
Therefore, the effective Hamiltonian governing the atomic $J=5/2$ Ce$^{3+}$
multiplet is
\begin{equation}
\hat{H}_{\mathrm{Ce}} = k_B T_{\mathrm{CEF}} \sum_{j=1}^2 \sum_{\sigma=\pm}
|\Gamma_8,j,\sigma\rangle\langle\Gamma_8,j,\sigma|
+ \, g \mu_B \mathbf{J} \cdot \mathbf{H} ,
\end{equation}
where $\mu_B$ is the Bohr magneton and the Land\'{e} $g$ factor for
Ce$^{3+}$ ($J = 5/2$, $L = 3$, $S = \frac{1}{2}$) is $g = 6/7$.

For a given $\mathbf{H}$, we diagonalized $\hat{H}_{\mathrm{Ce}}$ and
found the splitting $\Delta E$ between the lowest two energy eigenvalues.
We then used the relation $\Delta E = g_i \mu_B H$ to deduce an effective
value of $g_i(\mathbf{H})$ to insert into the $S=\thalf$ Kondo impurity model.
This effective $g$ factor is dependent on both the magnitude of $\mathbf{H}$
and its orientation relative to the crystal axes.
The strongest variation of $g_i$ with $H$ is found for fields oriented along
the $\langle100\rangle$ directions, while the weakest variation occurs for
fields along $\langle 111\rangle$.

To model the random orientation of our polycrystals relative to the field,
we averaged $g_i(\mathbf{H})$ over all directions of $\mathbf{H}$ at fixed
$H = |\mathbf{H}|$.
Table~\ref{table} shows the largest value, the smallest value, the mean,
and the standard deviation of $g_i$ for each field $H$ at which the specific
heat was measured, as well as the minimum splitting $\tilde{T}_{\mathrm{CEF}}$
between the second and third lowest eigenenergies.

The table shows that over the range of fields covered in our experiments,
the lowest-lying pair of states remains well separated in energy from the
remaining four states.
This separation ($\tilde{T}_{\mathrm{CEF}} \ge 55$\,K) justifies the
neglect of the higher levels at temperatures $T \le 10$\,K.

Second, it turns out that although the field causes quite strong mixing of CEF
levels, the distribution of $g_i$ values remains fairly narrow, and the mean
value shows a rather weak dependence on $H$.
We used just the mean value $\langle g_i\rangle$ in a NRG
calculation of $C_{\mathrm{imp}}$ at each magnetic field.
(The NRG calculations are computer intensive, and it was therefore impractical
to average over the entire distribution of $g_i$ values.)
We discuss the likely effect of this approximation in the next section.

\section{DISCUSSION}

Figure~\ref{fig:both_samples} shows the excess specific heat for
samples 1 and 2, at fields $H = 0$ and $H = 10$\,T.
At each field, the data for the two samples lie close to one another.
In particular, the location (in temperature) and height of the peak in
$\Delta C$ are very consistent between the samples.
Given that samples 1 and 2 were synthesized independently, the agreement
between their specific heats (both in zero and nonzero fields) suggests
that the data represent the intrinsic properties of
Ce$_{0.6}$La$_{0.4}$Pb$_3$, and are not merely sample-specific artifacts.

Figure~\ref{fig:compare} compares all the excess specific heat data for both
samples with the impurity specific-heat contribution $C_{\mathrm{imp}}$
computed as described in the preceding section.
Given that there is no adjustable parameter beyond the Kondo temperature
$T_K$ deduced from the zero field data, there is good agreement between
experiment and theory.
The Kondo model reproduces the field variation of the height of the
specific heat peak with very high accuracy.
The temperature of the peak in $\Delta C$ is also well described, the
only significant deviation occurring for $H=5$\,T.

In all cases $H>0$, the peak in the experimental data is somewhat broader than
predicted by the Kondo model, particularly on the low-temperature side of the
maximum in $\Delta C$.
This broadening may be partially attributable to the spread in $g_i$ values
arising from the distribution of angles between the magnetic field and the
cubic crystal axes.
As indicated by the maximum and minimum values listed in Table~\ref{table},
the $g_i$ distribution has longer tails on the low-$g_i$ side of
$\langle g_i \rangle$, which should tend to produce an asymmetry of the type
seen in $\Delta C$ vs $T$.

The rather minor differences between the field dependence observed in the
specific heat of Ce$_{0.6}$La$_{0.4}$Pb$_3$ and that predicted by the
single-impurity Kondo model suggest that there is no significant coherence
effect in the specific heat of this system, at least at temperatures
of order the single-ion Kondo scale and higher.
This provides a clear counter example to the two-fluid model
advanced\cite{Nakatsuji:04} for Ce$_x$La$_{1-x}$CoIn$_5$ and
Ce$_x$La$_{1-x}$IrIn$_5$, where coherence effects set in at temperatures
more than an order of magnitude above $T_K$.

\section*{ACKNOWLEDGMENTS}

This work was supported in part by the U.S.\ Department of Energy under Grant
No.\ DE-FG02-99ER45748 (R.P., C.R.R., and B.A.), by NSF under Grant
No.\ DMR-0312939 (K.I.), and by the University of Florida Physics REU Program
under NSF Grant No.\ DMR-0139579 (B.D.).




\end{document}